\documentclass[twocolumn,pra,preprintnumbers,showpacs,showkeys]{revtex4}
\usepackage{graphicx,amsmath,amsfonts,amssymb}
\usepackage{times}
\begin{document}
\title{Entanglement Sudden Death in Band Gaps}

\author{Ying-Jie Zhang}
\thanks{yingjiezhang2007@163.com}
\affiliation{Shandong Provincial Key Laboratory of Laser
Polarization and Information Technology, Department of Physics, Qufu
Normal University, Qufu 273165, Peoples Republic of China}

\date{\today}

\begin{abstract}
  Using the pseudomode method, we evaluate exactly time-dependent entanglement for two independent qubits, each coupled to a
 non-Markovian structured environment. Our results suggest a
 possible way to control entanglement sudden death by modifying the
 qubit-pseudomode detuning and the spectrum of the reservoirs.
 Particularly, in environments structured by a model of a
 density-of-states gap which has two poles, entanglement trapping
 and prevention of entanglement sudden death occur in the weak-coupling regime.
\end{abstract}



\maketitle

\section{ \textbf{Introduction}}
\indent Realistic quantum systems are affected by decoherence and
entanglement losses because of the unavoidable interaction with
their environments $[1]$. For example, in Markovian environments, in
spite of an exponential decay of the single-qubit coherence, the
entanglement between two qubits may completely disappear at a finite
time $[2,3]$. This phenomenon, known as entanglement sudden death
and experimentally proven to occur $[4,5]$, limits the time when
entanglement can be exploited for practical purposes. So the issue
of how to avoid or control entanglement sudden death-type
decorrelation in a realistic physical system is especially
important. Up to now, several ways were proposed for keeping the
atomic entanglement for a long time by suppressing spontaneous
emission. One way widely applied is to place the qubits in a
structured environment, say, microcavity $[6,7]$ or in the photonic
band gap of photonic crystals $[8]$. Other methods considered use
dynamic manipulation such as mode modulation $[9]$ or the quantum
Zeno effect $[10]$, which can be regarded as extensions
of the so-called bang-bang method $[11]$.\\
\indent In this paper we focus on the entanglement dynamics of two
qubits, each coupled to a non-Markovian structured environment. We
continue the investigation of physical effects that may control the
occurrence time of entanglement sudden death, and find that the
speed of disentanglement is closely related to qubit-pseudomode
detuning and the spectrum of the reservoirs. In particular, we
mainly consider two qubits respectively coupled to a bath of
oscillators with a density of states described by a frequency
dependent function $D(\omega)$. In some previous works $[12,13,14]$,
the spectrum of the structured reservoir is a Lorentzian, which only
give a single pseudomode to replace the effect of the structured
reservoir. Then the short-time behavior of the qubit system leads to
exponential decay $[15]$. However, if there are two or more poles
close to the real $\omega$ axis, it is not clear that a simple
picture of exponential decay will apply because the poles can
interfere with each other. So we focus on the structured reservoir
model of a band gap in which both Lorentzians are centered at the
same frequency, and the second is given a negative weighting in our
paper, and make comparison with the model of the function
$D(\omega)$ has one Lorentzian. \\
\section{ \textbf{Theoretical Model and Exact Dynamics of two qubits}}
\indent  Consider a model consisting of two qubits $A$ and
 $B$, each interacting with a common zero-temperature bosonic
 reservoir, denoted $a$ and $b$, respectively. We assume that each qubit-reservoir system
 is isolated and the reservoirs are initially in the vacuum
 state while two qubits are initially in an entangled state. Since each qubit evolves independently, we can learn how to
characterize the evolution of the overall system from the
qubit-reservoir dynamics. The interaction between a qubit and an
$N$-mode reservoir is described through the Hamiltonian (under the
rotating-wave approximation and setting $\hbar=1$)
\begin{equation}
\hat{H}_{j}=\omega_{j}\hat{\sigma}^{j}_{+}\hat{\sigma}^{j}_{-}+\sum_{k=1}^{N}\omega_{k}\hat{b}^{\dag}_{k}\hat{b}_{k}+\sum_{k=1}^{N}g_{k}(\hat{\sigma}^{j}_{-}\hat{b}^{\dag}_{k}+\hat{\sigma}^{j}_{+}\hat{b}_{k}),\label{01}
\end{equation}
where $\hat{b}^{\dag}_{k}$, $\hat{b}_{k}$ are the creation and
annihilation operators of quanta of the reservoir ($a$ or $b$),
$\hat{\sigma}^{j}_{+}=|1_{j}{\rangle}{\langle}0_{j}|$,
$\hat{\sigma}^{j}_{-}=|0_{j}{\rangle}{\langle}1_{j}|$ and
$\omega_{j}$ are the inversion operators and transition frequency of
the $j$-th qubit (j=$A$, $B$ and here
$\omega_{A}=\omega_{B}=\omega_{0}$ ); $\omega_{k}$ and $g_{k}$ are
the frequency of the mode $k$
of the reservoir and its coupling strength with the $j$-th qubit.\\
\indent Let us consider the case when the j-th qubit is initially in
the excited state and its corresponding reservoir is in the vacuum
state. Using the pseudomode method, we focus on an idealized model
of a band gap (or photon density of states gap) in which both
Lorentzians are centered at the same frequency, and the second is
given a negative weighting, so that $[16]$
\begin{equation}
D(\omega)=\frac{W_{1}\Gamma_{1}}{(\omega-\omega_{c})^{2}+(\frac{\Gamma_{1}}{2})^{2}}-\frac{W_{2}\Gamma_{2}}{(\omega-\omega_{c})^{2}+(\frac{\Gamma_{2}}{2})^{2}},\label{02}
\end{equation}
where now the weights of the two Lorentzians are such that
$W_{1}-W_{2}=1$ and $\Gamma_{2}<\Gamma_{1}$ ensure positivity of
$D$. $\omega_{c}$ is the center of the spectrum, and $\Gamma_{1}$,
$\Gamma_{2}$ are the full width at half maximum of two Lorentzians,
respectively. The effect of the Lorentzian with negative weight is
to introduce a dip into the density of states function $D(\omega)$
where the coupling of the qubit will be inhibited $[17,18]$. The two
poles in $D$ are located at $\omega_{c}-i\Gamma_{1}/2$ and
$\omega_{c}-i\Gamma_{2}/2$, and there is a change in sign of the
residues of $D$ between these poles. The exact pseudomode master
equation associated with the density of states gap in
Eq.$(\ref{02})$ is given by
\begin{eqnarray}
\frac{d\rho}{dt}=&-&i[H_{0}^{j},\rho]-\frac{\Gamma'_{1}}{2}[a^{\dag}_{1}a_{1}\rho-2a_{1}{\rho}a^{\dag}_{1}+{\rho}a^{\dag}_{1}a_{1}]\nonumber\\
&-&\frac{\Gamma'_{2}}{2}[a^{\dag}_{2}a_{2}\rho-2a_{2}{\rho}a^{\dag}_{2}+{\rho}a^{\dag}_{2}a_{2}],\label{03}
\end{eqnarray}
where
\begin{eqnarray}
H_{0}^{j}&=&\omega_{0}\sigma^{j}_{+}\sigma^{j}_{-}+\omega_{c}a^{\dag}_{1}a_{1}+\omega_{c}a^{\dag}_{2}a_{2}+\Omega_{0}[a^{\dag}_{2}\sigma^{j}_{-}+a_{2}\sigma^{j}_{+}]\nonumber\\
&+&V(a^{\dag}_{1}a_{2}+a_{1}a^{\dag}_{2}), \label{04}
\end{eqnarray}
where $\rho$ is the density operator for the j-th qubit and the
pseudomodes of its corresponding reservoir; and $a_{1}$ and $a_{2}$
are the annihilation operators of the two pseudomodes decaying with
deacy rates $\Gamma'_{1}=W_{1}\Gamma_{2}-W_{2}\Gamma_{1}$ and
$\Gamma'_{2}=W_{1}\Gamma_{1}-W_{2}\Gamma_{2}$ respectively. The two
pseudomodes are coupled and
$V=\sqrt{W_{1}W_{2}}(\Gamma_{1}-\Gamma_{2})/2$ is the strength of
the coupling. The qubit interacts coherently with the second
pseudomode (the strength of the coupling $\Omega_{0}$), which is in
turn coupled to the first one. Both pseudomodes are leaking into
independent Markovian reservoirs. The set of ordinary differential
equations associated to the master equation $(\ref{03})$ is
\begin{eqnarray}
i\frac{dc_{1}}{dt}&=&\omega_{0}c_{1}+\Omega_{0}b_{2},\nonumber\\
i\frac{db_{1}}{dt}&=&z'_{1}b_{1}+Vb_{2},\nonumber\\
i\frac{db_{2}}{dt}&=&z'_{2}b_{2}+Vb_{1}+\Omega_{0}c_{1}, \label{05}
\end{eqnarray}
where $c_{1}$, $b_{1}$, and $b_{2}$ are the complex amplitudes for
the states with one excitation in the qubit, one excitation in the
first pseudomode, and one excitation in the second pseudomode,
respectively. The positions of the true poles are
$z'_{1}=\omega_{c}-i\Gamma'_{1}/2$ and
$z'_{2}=\omega_{c}-i\Gamma'_{2}/2$. Due to the initial state of the
j-th qubit and its corresponding reservoir is
$|1\rangle_{j}\otimes|\bar{0}\rangle$ (with the state
$|\bar{0}\rangle=\prod_{k=1}^{N}|0_{k}\rangle$), then $c_{1}(0)=1$,
and $b_{1}(0)=b_{2}(0)=0$, so we can acquire the exact solutions
($c_{1}$, $b_{1}$ and $b_{2}$) of the differential equations
(\ref{05}) easily through a computer program.\\
\indent According to the differential equations (\ref{05}), for an
initial state of the form
\begin{equation}
|\Psi(0)\rangle=(\alpha|00\rangle_{AB}+\beta|11\rangle_{AB})\otimes|\overline{0}\rangle_{a}|\overline{0}\rangle_{b},\label{06}
\end{equation}
where $\alpha$, $\beta$ are real, then the time evolution of the
total system is (using the above pseudomode method)
\begin{eqnarray}
|\Psi(t)\rangle&=&\beta(c_{1}|1\rangle_{A}|0_{1}0_{2}\rangle_{a}+b_{1}|0\rangle_{A}|1_{1}0_{2}\rangle_{a}+b_{2}|0\rangle_{A}|0_{1}1_{2}\rangle_{a})\nonumber\\
&\times&(c_{1}|1\rangle_{B}|0_{1}0_{2}\rangle_{b}+b_{1}|0\rangle_{B}|1_{1}0_{2}\rangle_{b}+b_{2}|0\rangle_{B}|0_{1}1_{2}\rangle_{b})\nonumber\\
&+&\alpha|00\rangle_{AB}|0_{1}0_{2}\rangle_{a}|0_{1}0_{2}\rangle_{b},\label{07}
\end{eqnarray}
with $|0_{1}0_{2}\rangle_{a/b}$, $|1_{1}0_{2}\rangle_{a/b}$ and
$|0_{1}1_{2}\rangle_{a/b}$ mean the structured reservoir $a$ or $b$
states with no excitation in two pseudomodes, only one excitation in
the first pseudomode, and one excitation in the second pseudomode,
respectively. Then we can determine the two qubits dynamics. In
particular, in the standard two qubits basis
$\mathcal{C}=\{|00\rangle_{AB}, |01\rangle_{AB}, |10\rangle_{AB},
|11\rangle_{AB}\}$, the reduced density matrix of the two qubits at
time $t$ result as
\begin{eqnarray}
\rho_{AB}(t)&=&(\alpha^{2}+\beta^{2}(1-|c_{1}|^{2})^{2})|00\rangle_{AB}\langle00|\nonumber\\
&+&(\beta^{2}|c_{1}|^{2}(1-|c_{1}|^{2}))|01\rangle_{AB}\langle01|\nonumber\\
&+&(\beta^{2}|c_{1}|^{2}(1-|c_{1}|^{2}))|10\rangle_{AB}\langle10|\nonumber\\
&+&\beta^{2}|c_{1}|^{4}|11\rangle_{AB}\langle11|\nonumber\\
&+&\alpha{\beta}c_{1}^{2}|00\rangle_{AB}\langle11|\nonumber\\
&+&\alpha{\beta}(c_{1}^{2})^{*}|11\rangle_{AB}\langle00|.\label{08}
\end{eqnarray}
We use the concurrence $C$ $[19]$, which attains its maximum value
$1$ for maximally entangled states and vanishes for separable
states, to analyze the two-qubit entanglement dynamics. For
$\rho_{AB}(t)$, its concurrence can be derived from $[19]$, as
\begin{equation}
C(t)=2\max\{0,
\alpha\beta|c_{1}|^{2}-\beta^{2}|c_{1}|^{2}(1-|c_{1}|^{2})\}.\label{09}
\end{equation}
\section{ \textbf{Numerical Results and Discussions}}
\indent Similar to the result in Ref.$[20]$, it is easy to find that
the two-qubit entanglement can occur sudden death for
$\alpha<\beta$. In this non-Markovian system-reservoir coupling
model, the concurrence of the two qubits will vanish forever at a
finite interval in weak-coupling regime, but for the strong-coupling
regime, there will be entanglement revival after entanglement sudden
death. In the following, we mainly investigate that the speed of the
occurrence of entanglement sudden death is related to the spectrum
of the reservoirs ($\Gamma_{2}$) and the
 qubit-pseudomode detuning ($\Delta=\omega_{c}-\omega_{0}$).\\
\indent First we study the relation between the entanglement sudden
death and  the spectrum of the reservoirs. Fig.$1$ shows the
entanglement dynamics of two qubits in the non-Markovian
weak-coupling regime with $\Gamma_{1}/2=10\Omega_{0}$. We compare
the entanglement dynamics of the two qubits for three different
values of the width of the second Lorentzian spectral function,
namely, $\Gamma_{2}/2=\Omega_{0}$, $2\Omega_{0}$, $9\Omega_{0}$. As
in Fig.$1(a)$, the qubits are on resonance with the center of the
spectrum, $\Delta=0$. The concurrence decreases monotonically down
to zero in a finite time. It is interesting to find that the speed
of occurrence of entanglement sudden death can increase with
$\Gamma_{2}$ increasing. Similar behavior happens when two qubits
are near resonance with the center of the spectrum. However, When
the qubits are far off-resonant with the center of the spectrum,
$\Delta\gg\Omega_{0}$, as shown in Fig.$1(b)$, where we choose
$\Delta=10\Omega_{0}$, the speed of occurrence of entanglement
sudden death decreases as $\Gamma_{2}$ increases. We can give an
intuitive explanation for these results. As we can see in Fig.$2$,
the density of the spectrum $D(\omega)$ increases monotonically as
$\Gamma_{2}$ increases for $\Delta=0$, while the density of the
spectrum $D(\omega)$ decreases monotonically as $\Gamma_{2}$
increases when $\Delta=10\Omega_{0}$. It is proved that the
entanglement sudden death is determined by the modes of the spectrum
which are on resonance with the qubits: the speed of the occurrence
of entanglement sudden death decreases (increases) as the density of
these modes decreases (increases).\\
\begin{figure}
\includegraphics[scale=1.1]{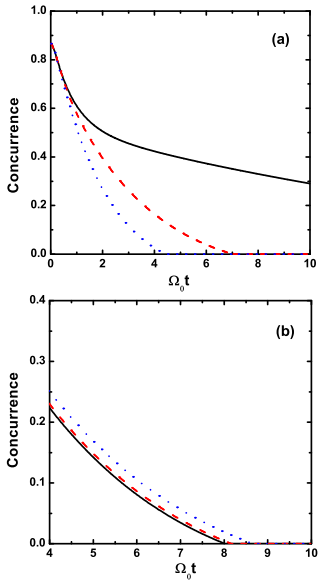}
\caption{\label{fig1}Time evolution of the concurrence of two qubits
as a function of the dimensionless quantity $\Omega_{0}t$ in
non-Markovian weak-coupling regime, with (a) $\Delta=0$ and (b)
$\Delta=10\Omega_{0}$. For the cases of (i)
$\Gamma_{2}/2=\Omega_{0}$ (solid dark curve), (ii)
$\Gamma_{2}/2=2\Omega_{0}$ (dashed red curve), (iii)
$\Gamma_{2}/2=9\Omega_{0}$ (dotted blue curve). The parameters used
are: $\Gamma_{1}/2=10\Omega_{0}$, $W_{1}=1.1$, $W_{2}=0.1$ and
$\alpha=\frac{1}{2},\beta=\frac{\sqrt{3}}{2}$.}
\end{figure}
\begin{figure}
\includegraphics[scale=1.2]{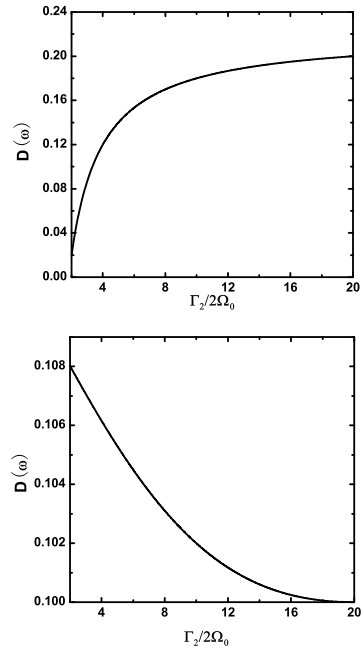}
\caption{\label{fig2}The density of the spectrum $D(\omega)$ as a
function of the dimensionless quantity $\Gamma_{2}/2$ in
non-Markovian weak-coupling regime, with (a) $\Delta=0$ and (b)
$\Delta=10\Omega_{0}$. The parameters used are:
$\Gamma_{1}/2=10\Omega_{0}$, $W_{1}=1.1$, $W_{2}=0.1$.}
\end{figure}
\begin{figure}
\includegraphics[scale=0.6]{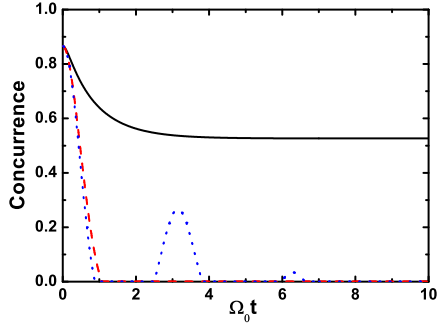}
\caption{\label{fig3}Time evolution of the concurrence of two qubits
as a function of the dimensionless quantity $\Omega_{0}t$, for the
cases of (i)weak-coupling regime
$\Gamma_{1}/2=11\Omega_{0},\Gamma_{2}/2=\Omega_{0}$ (solid dark
curve), (ii)intermediate-coupling regime
$\Gamma_{1}/2=1.1\Omega_{0},\Gamma_{2}/2=0.1\Omega_{0}$ (dashed red
curve), (iii)strong-coupling regime
$\Gamma_{1}/2=0.11\Omega_{0},\Gamma_{2}/2=0.01\Omega_{0}$ (dotted
blue curve). The parameters used are: $W_{1}=1.1$, $W_{2}=0.1$ and
$\alpha=\frac{1}{2},\beta=\frac{\sqrt{3}}{2}$.}
\end{figure}
\indent If the structured reservoir only contains a Lorentzian
($\Gamma_{2}=0$), from the Ref.$[12]$, we note that the speed of
disentanglement decreases as the width of the Lorentzian spectral
($\Gamma_{1}$ is replaced by $\lambda$ in Ref.$[12]$) increases
on/near the resonant couplings, and the speed of disentanglement
increases as $\lambda$ increases for large deduning coupling.
However, in our paper the density of the spectrum $D(\omega)$ is
composed by two Lorentzians, and the second is given a negative
weighting. Then the counter results can be acquired when we consider
the width of the second Lorentzian spectral $\Gamma_{2}$ influence
the speed of the occurrence of entanglement sudden death. That is to
say, $\Gamma_{1}$ and $\Gamma_{2}$ have the opposite effects on the
speed of the occurrence of entanglement sudden death. So we can keep
two-qubit entanglement for a long time trough choosing suitable
spectrum of the reservoirs.\\
\indent As we know, the effect of the Lorentzian with negative
weight is to introduce a dip into the density of states function
$D(\omega)$. For a perfect gap, where $D(\omega_{c})=0$, we would
also have $W_{1}/\Gamma_{1}=W_{2}/\Gamma_{2}$, then the single-qubit
excited-state population trapping can occurs $[15]$. According to
Ref.$[21]$, we can find that there is a direct link between the
time-dependent entanglement and single-qubit excited-state
population for independent qubits, each coupled to a
zero-temperature bosonic environment. So there also will appear
two-qubit entanglement trapping if the qubits are resonant with the
gap in our paper. But it is surprising to see that the entanglement
trapping only can occur in the weak-coupling regime, and
entanglement sudden death still appears quickly in the
intermediate-coupling regime and strong-coupling regime. For the
intermediate-coupling regime, the two-qubit entanglement can vanish
forever. Due to the strong non-Markovian effects, the entanglement
of the two qubits can arise the phenomenon of entanglement sudden
death and revival in the strong-coupling regime (as shown in
Fig.$3$).\\
\begin{figure}
\includegraphics[scale=0.6]{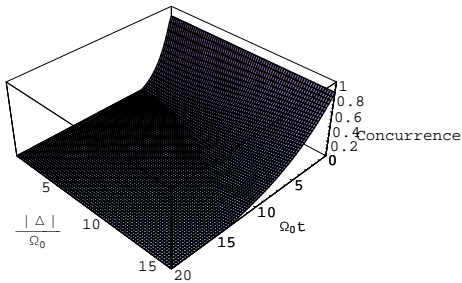}
\caption{\label{fig4}Time evolution of the concurrence of two qubits
as a function of the dimensionless quantity $\Omega_{0}t$ and
$|\Delta|$ in reservoirs structured by a model of one Lorentzian
spectral density. The parameters used are: $W_{1}=1$,
$\Gamma_{1}/2=10\Omega_{0}$ and
$\alpha=\frac{1}{2},\beta=\frac{\sqrt{3}}{2}$.}
\end{figure}
\begin{figure}
\includegraphics[scale=0.6]{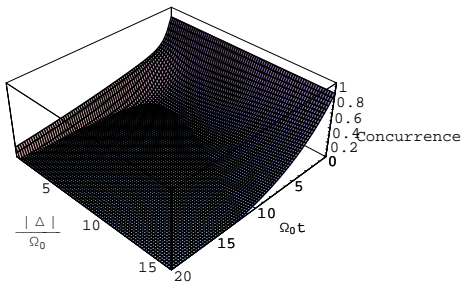}
\caption{\label{fig5}Time evolution of the concurrence of two qubits
as a function of the dimensionless quantity $\Omega_{0}t$ and
$|\Delta|$ in reservoirs structured by a model of a
density-of-states gap which has two Lorentzians. The parameters used
are: $W_{1}=1.1$, $W_{2}=0.1$, $\Gamma_{1}/2=10\Omega_{0}$,
$\Gamma_{2}/2=\Omega_{0}$, and
$\alpha=\frac{1}{2},\beta=\frac{\sqrt{3}}{2}$.}
\end{figure}

 \indent Then we fix the spectrum of the reservoirs to study
the relation between the entanglement sudden death and
qubit-pseudomode detuning $\Delta$. In Fig.$4$ we show the
entanglement dynamics for two qubits in a structured reservoir,
which only has one Lorentzian spectral. Fig.$5$ shows the
entanglement dynamics when the qubits interact with two independent
reservoirs structured by a model of a density-of-states gap which
has two Lorentzians. A comparison between Fig.$4$ and Fig.$5$
reveals that, both for the weak-coupling regime, the speed of the
occurrence of entanglement sudden death only can decrease as
$|\Delta|$ increases in the one Lorentzian model. However, there
exist a critical value $|\Delta_{c}|$ (with
$\Delta_{c}^{2}=\frac{\Gamma_{1}^{2}\sqrt{W_{2}\Gamma_{2}}-\Gamma_{2}^{2}\sqrt{W_{1}\Gamma_{1}}}{4\sqrt{W_{1}}(\sqrt{\Gamma_{1}}-\sqrt{\Gamma_{2}})}$)
in the band gap model, then the speed of occurrence of entanglement
sudden death will increase as $|\Delta|$ increases when
$|\Delta|<|\Delta_{c}|$, and entanglement sudden death can occur
more slowly with $|\Delta|$ increasing at $|\Delta|>|\Delta_{c}|$. A
physical interpretation of the result is that the density of the
spectrum $D(\omega)$ decreases monotonically as $|\Delta|$ increases
in the one Lorentzian model, while in the band gap model,
$D(\omega)$ increases as $|\Delta|$ increases when
$|\Delta|<|\Delta_{c}|$ and decreases as $|\Delta|$ increases at
$|\Delta|>|\Delta_{c}|$. The relation between $D(\omega)$ and
$\Delta$ is clearly shown in Fig.$6$, for example, when
$\Gamma_{1}/2=10\Omega_{0},\Gamma_{2}/2=\Omega_{0}$, we can
calculate $\Delta_{c}=\pm3.53\Omega_{0}$.
\begin{figure}
\includegraphics[scale=0.6]{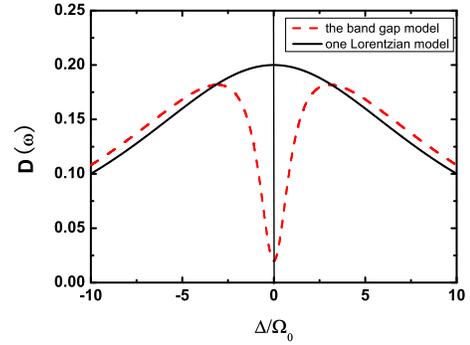}
\caption{\label{fig6}The density of the spectrum $D(\omega)$ as a
function of the dimensionless quantity $\Delta$ in non-Markovian
weak-coupling regime, for the one Lorentzian model
$\Gamma_{1}/2=10\Omega_{0}$, $W_{1}=1$ (solid dark curve) and the
band gap model $\Gamma_{1}/2=10\Omega_{0}$,
$\Gamma_{2}/2=\Omega_{0}$, $W_{1}=1.1$ and $W_{2}=0.1$ (dashed red
curve).}
\end{figure}
\section{ \textbf{Conclusion}}
\indent In summary, we have presented a non-Markovian model
describing the exact entanglement dynamics of two qubits, each
interacted with a structured reservoir. We have brought to light new
relations between the speed of the occurrence of entanglement sudden
death and the spectrum of the reservoirs/the
 qubit-pseudomode detuning for qubits prepared in entangled state. We firstly find that the
speed of occurrence of entanglement sudden death is a increasing
(decreasing) function of the width of the second Lorentzian spectral
$\Gamma_{2}$ when the qubits are on/near resonance (large detuning)
with the pseudomodes of the reservoirs. Due to the density of the
spectrum $D(\omega)$ is composed by two Lorentzians in our paper,
and the second is given a negative weighting, it is interesting to
find that $\Gamma_{1}$ and $\Gamma_{2}$ have the opposite effects on
the speed of the occurrence of entanglement sudden death. Then for a
perfect gap, where $D(\omega_{c})=0$, we would also have
$W_{1}/\Gamma_{1}=W_{2}/\Gamma_{2}$, the qubits' entanglement
trapping and prevention of entanglement sudden death can occur in
the weak-coupling regime, but entanglement sudden death still
appears quickly in the intermediate-coupling regime and
strong-coupling regime. Next, a comparison between the one
Lorentzian model and the band gap model, we find the speed of the
occurrence of entanglement sudden death only can decrease as
$|\Delta|$ increases in the one Lorentzian model and a critical
value $|\Delta_{c}|$ exist in the band gap model. If
$|\Delta|<|\Delta_{c}|$, the speed of occurrence of entanglement
sudden death will increase as $|\Delta|$ increases, and when
$|\Delta|>|\Delta_{c}|$, entanglement sudden death can occur more
slowly with $|\Delta|$ increasing. The results here obtained
evidence the entanglement can be preserved or controlled by
modifying the spectrum of the environment and highlight
the potential of reservoir engineering for controlling and manipulating the dynamics of quantum systems.\\
\indent Our results would apply to cavity QED experiments with
trapping ions, and to circuit QED experiments. Entanglement between
two remotely located trapped atomic ions has been recently
demonstrated $[22]$ and multiparticle-entangled states can be
generated and fully characterized via state tomography $[23]$.
Moreover, field coupling and coherent quantum state storage between
two Josephson phase qubits has been achieved through a microwave
cavity on chip $[24]$. Due to the possibilities for realizing strong
coupling conditions between atoms and a high finesse cavity $[25]$,
a deep understanding of the non-Markovian dynamics is now
indispensable.
\section{ \textbf{Acknowledgments}}
This work is supported by National Natural Science Foundation of
China under Grant No. 10774088 and the key Program of Science
Foundation of China under Grant No. 10534030.

\end{document}